\documentclass[a4paper,onecolumn,11pt,accepted=2023-01-18]{quantumarticle}
\pdfoutput=1 
 
\usepackage{authblk}
\usepackage{fullpage, xcolor}
\usepackage{amssymb}
\usepackage{amsmath}
\usepackage{amsfonts}
\usepackage{amsthm}
\usepackage{dsfont}
\usepackage{comment}
\usepackage{enumerate}
\usepackage{float}
\usepackage{indentfirst}
\usepackage{booktabs}
\usepackage{array}
\usepackage{colortbl}
\usepackage{subcaption}
\usepackage{graphicx}
\usepackage{listings}
\usepackage[colorlinks,linkcolor=blue,anchorcolor=blue,citecolor=blue]{hyperref}
\usepackage{cleveref}
\usepackage[english]{babel}
\usepackage[T1]{fontenc}
\usepackage{hyperref}

\newcommand{\ynote}[1]{{{\color{purple}({\bf Yupan:} #1)}{}}}

\usepackage{lipsum}

\newcommand{\problem}[1]{\ensuremath{\mathsf{#1}\xspace}} \newcommand\defproblem[5]{
\begin{definition}[#1]\label{#2}
#3 \\
\hspace{1cm} \textbf{Yes.} #4 \\
\noindent\;\;\textbf{No.} #5
\end{definition} 
}

\newcommand{\complex}{\mathbb{C}}

\usepackage[ruled,vlined]{algorithm2e}

\usepackage{enumerate}
\usepackage{framed}
\usepackage[title]{appendix}
\usepackage{colortbl}
\usepackage{comment}
\usepackage[modulo]{lineno}
\usepackage{cite}

\allowdisplaybreaks[4]

\begin{document}

\newtheorem{theorem}{Theorem}[section]
\newtheorem{definition}[theorem]{Definition}
\newtheorem{lemma}[theorem]{Lemma}
\newtheorem{proposition}[theorem]{Proposition}
\newtheorem{claim}[theorem]{Claim}
\newtheorem{corollary}[theorem]{Corollary}
\newtheorem{conjecture}[theorem]{Conjecture}
\newtheorem{remark}[theorem]{Remark}

\newcommand{\class}[1]{\ensuremath{\mathsf{#1}}}
\newcommand{\Pc}{\class{P}}
\newcommand{\BQP}{\mathsf{BQP}}
\newcommand{\BPP}{\class{BPP}}
\newcommand{\NP}{\class{NP}}
\newcommand{\QMA}{\class{QMA}}
\newcommand{\MA}{\class{MA}}
\newcommand{\AM}{\class{AM}}
\newcommand{\StoqMA}{\class{StoqMA}}
\newcommand{\LHP}{\mathsf{LHP}}
\newcommand{\SAT}{\mathsf{SAT}}
\newcommand{\PSPACE}{\mathsf{PSPACE}}
\newcommand{\bra}[1]{\langle #1|}
\newcommand{\ket}[1]{| #1 \rangle}
\newcommand{\braket}[3]{\langle #1 | #2 | #3 \rangle}
\newcommand{\ketbra}[2]{| #1 \rangle \langle #2 |}
\newcommand{\kb}[1]{| #1 \rangle \langle #1 |}
\newcommand{\innerprod}[2]{\langle #1 | #2 \rangle}
\newcommand{\expval}[1]{\langle #1 \rangle}
\newcommand{\norm}[1]{\left\| #1 \right\|}
\newcommand{\Tr}{\mathrm{Tr}}

\newcommand{\cH}{\mathcal{H}}
\newcommand{\cL}{\mathcal{L}}
\newcommand{\poly}{\mathrm{poly}}
\newcommand{\negl}{\mathrm{negl}}
\newcommand{\setunsat}{\mathrm{set\text{-}unsat}}
\newcommand{\SetCSP}{\mathrm{SetCSP}}
\newcommand{\eps}{\varepsilon}

\renewcommand{\Pr}[1]{\mathrm{Pr}\left[#1\right]}
\newcommand{\E}{\mathop{\mathbb{E}}}
\newcommand{\h}{\mathrm{H}}
\newcommand{\s}{\mathrm{S}}

\newcommand{\PSLH}{\problem{ProjUSLH}}
\newcommand{\binset}{\{0,1\}}
\newcommand{\calL}{\mathcal{L}}

\title{$\StoqMA$ vs. $\MA$: the power of error reduction}

\author{Dorit Aharonov}
\affiliation{School of Computer Science and Engineering, The Hebrew University of Jerusalem}
\affiliation{Qedma Quantum Computing Ltd.}
\email{doria@cs.huji.ac.il}

\author{Alex B. Grilo}
\affiliation{Sorbonne Université, CNRS, LIP6}
\email{Alex.Bredariol-Grilo@lip6.fr}

\author{Yupan Liu}
\affiliation{Graduate School of Mathematics, Nagoya University}
\email{yupan.liu@gmail.com}
%yupan.liu.e6@math.nagoya-u.ac.jp

\maketitle
\begin{abstract}
\StoqMA{} characterizes the computational hardness of stoquastic local Hamiltonians, which is a family of Hamiltonians that does not suffer from the sign problem. 
Although error reduction is commonplace for many complexity classes, such as \class{BPP}, \class{BQP}, \MA{}, \QMA{}, etc., this property remains open for \StoqMA{} since Bravyi, Bessen and Terhal defined this class in 2006. 
In this note, we show that error reduction for $\StoqMA$ will imply that \StoqMA{} $=$ \MA{}. 
\end{abstract} \section{Introduction}\label{sec:intro}
The local Hamiltonian problem is a cornerstone of quantum complexity theory. The input to the problem consists of a $k$ -local Hamiltonian $H = \frac{1}{m} \sum_{i \in [m]} H_i$, where each term $H_i$ acts non-trivially on $k$-qubits out of an $n$ qubit system\footnote{A classical local Hamiltonian could be interpreted as an instance of the Constraint Satisfaction Problem (CSP) problem. In particular, each a local term in Hamiltonian matches a constraint, where variables in the CSP instance correspond to qubits. Moreover, Hamiltonian's ground state is associated with an assignment, and the ground-energy is precisely the maximal violation of such an assignment. }, as well as two numbers, $\alpha$ and $\beta$. We are asked then to decide if the ground-energy of $H$ is below $\alpha$ or above $\beta$, when we are promised that one of the cases hold. 
This problem is a computational version of the fundamental problem in condensed matter physics, of understanding the ground-energy of many body quantum systems. 
Kitaev~\cite{KSV02} showed that for some $\alpha$ and $\beta$ such that $\beta - \alpha \geq 1/\poly(n)$, the $k$-local Hamiltonian problem is complete for \QMA{}, the quantum analog of \NP{}. 
Many researchers have studied the complexity of the local Hamiltonian problem for restricted families of local Hamiltonians, closer to physical models, e.g., \cite{KR03,KKR06,OT08,AGIK09,CGW14,CM16}. 

In this work we are interested in a particular subfamily of Hamiltonians called {\em stoquastic Hamiltonians}.\footnote{We avoid giving the formal definition of these Hamiltonians here and leave it to Section \ref{subsec:stoq-LH}.}
Roughly, stoquastic Hamiltonians are those Hamiltonians whose off-diagonal terms are non-positive. From a physics point of view, stoquastic Hamiltonians arise naturally since they do not exhibit what is known as the {\em sign problem}, making them more amenable to classical simulations by Monte Carlo methods. Moreover, the first adiabatic algorithms were proposed with stoquastic Hamiltonians~\cite{Farhi}, and the first models of D-Wave quantum devices were implemented in this model~\cite{JAG+11,AL18}.  

Several works, led by Bravyi and Terhal~\cite{BDOT06,BBT06,BT10}, revealed  the important role that stoquastic Hamiltonians play in quantum computational complexity. \cite{BBT06,BT10} showed that deciding whether a stoquastic Hamiltonian is {\it frustration-free}, namely (with some normalization) has ground-energy $0$, or its ground-energy is at least $1/\poly(n)$, is \MA{}-complete. This is surprising, since it is the {\em first} natural \MA{}-complete problem -- and it is stated using quantum language.\footnote{More precisely, \cite{BT10} show that such a problem is \MA{}-hard and lies in $\class{Promise}\MA$. For simplicity, we hereafter call such problems $\MA$-complete.} Moreover, it was proven by  \cite{BBT06} that if we are interested in deciding if a stoquastic Hamiltonian has ground-energy at most $\alpha$ or at least $\alpha + 1/\poly(n)$, for some more general non-zero $\alpha$, then the problem becomes \StoqMA{}-complete, where \StoqMA{} is defined to be a (somewhat artificial) generalization of \MA{}, which sits between \MA{} and \QMA{}.

The class \StoqMA{} is our focal point in this paper;
Despite its unnatural definition, which we will review in a moment, it plays a prominent role in Hamiltonian complexity.
In particular, it was shown in  \cite{CM16,BH17} that it participates in a very interesting classification theorem about quantum local Hamiltonians: for any type of local Hamiltonians,\footnote{By type of local Hamiltonians, we mean that the local Hamiltonian has special properties. For example, one of the types considered in~\cite{CM16} consists of local Hamiltonians for which there is a tensor product of one-qubit unitaries that diagonalizes all local terms simultaneously.} the complexity class of deciding if the ground-energy of Hamiltonians of this type is at most $\alpha$ or at least $\alpha + 1/\poly(n)$, falls in one of four categories: it is either in \class{P}, \NP{}-complete,  \class{StoqMA}-complete or \QMA{}-complete.
To put this classification in context, recall Schaefer's dichotomy theorem~\cite{Schaefer} which is the analog classification theorem for Max-k-SAT. 
Schaefer's dichotomy theorem states that when we consider classical Hamiltonians (i.e. the Max-k-SAT problem), the problems are either in \class{P} or $\NP$-complete.
Since Max-k-SAT is exactly a special case of local Hamiltonian problems, it would have been natural to expect that the classification in the quantum case would be to {\it three} classes, with the only new addition to \class{P} and \NP{} being the class \QMA{}. Yet, surprisingly, \cite{CM16} shows that there is a fourth class which  also participates in this classification, 
namely, \StoqMA{}.

We now devote some time to define \StoqMA{} informally (See \Cref{sec:complexity-classes} for a formal definition). As in standard proof systems (such as \NP, \QMA, etc.), in \StoqMA{} we consider a protocol for some problem~\footnote{For us, we consider a problem as a set of two disjoint sets $\mathcal{L}_{yes}$, known as positive instances and $\mathcal{L}_{no}$, known as negative instances. The verifier is then given $x \in \mathcal{L}_{yes} \cup \mathcal{L}_{no}$, and her goal is to decide if $x \in \mathcal{L}_{yes}$ or if $x \in \mathcal{L}_{no}$.} involving two parties, a computationally unbounded prover and a computationally bounded (usually polynomial time) verifier. In this protocol, the prover sends a quantum state to the verifier.   Then the verifier of the \StoqMA{} protocol runs her verification algorithm that has a very particular structure: First, the verifier's computation is given by a classical reversible circuit, viewed as a quantum circuit (i.e., it is performed coherently). Secondly, its input consists of three parts: $i)$ the quantum proof provided by the (unbounded) prover; $ii)$ auxiliary qubits in the state $\ket{0}$; and $iii)$ auxiliary qubits in the state $\ket{+} =  (\ket{0} + \ket{1}) / \sqrt{2}$.
After applying the classical reversible circuit, a certain qubit designated as the output qubit is measured in the $\ket{+},\ket{-}$ basis, and the verifier accepts iff the output is $\ket{+}$. 
A problem is then said to be in the complexity class $\StoqMA(a,b)$, for some $a>b\ge 0$,\footnote{We can assume that the soundness parameter $b$ is at least $\frac{1}{2}$, WLOG, as argued in \Cref{rem:soundness-bound}. } if for yes-instances, there is a quantum state that the prover can send that makes the verifier accept with probability at least $a$, whereas for no-instances, any quantum state provided by the prover makes the verifier accept with probability at most $b$.

As in other complexity classes with non-perfect completeness and soundness parameters, we would like to understand 
how does
$\StoqMA(a,b)$ depend on 
the parameters $a,b$.
First, it is very easy to see that $\StoqMA(a,b)\subseteq \StoqMA(a',b')$ if $a'\le a$ and $b'\ge b$.  For several complexity classes such as \BPP{}, \class{BQP}, \MA{}, \QMA{}, etc., much more can be said about the dependence on parameters: the class stays the same as long as the completeness and soundness differ by at least some inverse polynomial. This is done by what is called 
{\it error reduction}, where we repeat the verification in parallel and do majority vote on the parallel output qubits. Peculiarly, it is not known how to do error reduction for \StoqMA{}! The standard approach does not seem to work for $\StoqMA$ (at least not in a straightforward way) because the majority vote between the output of the parallel runs needs to be done in the Hadamard basis, which is not (known to be) possible using a classical circuit, as required in \StoqMA{}.\footnote{See \Cref{rem:straightforward-error-reduction} for more details.} Yet, we conjecture here:

\begin{conjecture}[Error reduction for $\StoqMA$]
	\label{conj:StoqMA-error-reduction}
	For any $a,b$ such that $1/2 \leq b < a \leq 1$ and $a-b \geq 1/\poly(n)$, 
	\[\StoqMA(a,b) \subseteq \StoqMA\left(1-2^{-l(n)},\frac{1}{2}+2^{-l(n)}\right),\]
	where $l(n)$ is a polynomial of $n$. 
\end{conjecture}

If this conjecture holds, we could write 
\StoqMA{} without specifying the parameters (as is done for example in \BPP{} and \class{BQP}); essentially that would mean $\StoqMA(a,b)$ with any $a,b$ such that $a-b>1/\poly(n)$. 

The main result in this paper is that surprisingly, if this natural property of error reduction holds for
\StoqMA{}, this would imply a very interesting conclusion: that $\StoqMA{}$ is equal to \MA{}. In fact, we show this already follows from a much weaker version of error reduction. 
We state this implication as another conjecture: 

\begin{conjecture}\label{conj:stoqma-equals-ma}
For any $a,b$ such that  $1/2 \leq b < a \leq 1$ and $a - b \geq 1/\poly(n)$, we have that \StoqMA{}$\left(a,b\right)$ = \MA{}.
\end{conjecture}

Before stating our result connecting Conjectures \ref{conj:StoqMA-error-reduction} and \ref{conj:stoqma-equals-ma} more formally, 
let us try to argue why Conjecture \ref{conj:stoqma-equals-ma} is a natural conjecture, independently. To this end, let us compare \StoqMA{} to other better understood complexity classes. 
First, \cite{BBT06} showed that  
\StoqMA{}$(a,b)$ is contained in \AM{} for every $a - b \geq 1/\poly(n)$\footnote{Recall \cite{Babai85} that \AM{} is a two-message randomized generalization of \NP{}, in which the message from the verifier consists of random bits}. Notice that this puts a strong upper bound on $\StoqMA{}$, when compared with the belief that $\BQP{}$, the class of problems solved in quantum polynomial-time, is not included in the polynomial hierarchy~\cite{RazT19}.

From the other side, it follows that \StoqMA{}($1,\frac{1}{2}+\negl(n)$) contains \MA{}. To see this, let us use the fact we can consider \MA{} with perfect completeness and $\negl(n)$ soundness error WLOG. If we consider the coherent version of \MA{} verifier, we can modify it to perform a controlled-SWAP where the control wire is the \MA{} output wire, and it swaps an auxiliary qubit $\ket{0}$  and an  auxiliary qubit $\ket{+}$ and uses the first of these registers as the $\StoqMA$ output. For completeness, the SWAP is always performed (due to perfect completeness for \MA{}) so we also have perfect completeness in the \StoqMA{} verifier. For no-instances, with $1-\negl(n)$ probability the $\MA$ verifier's output is $0$ and therefore there is no SWAP between $\ket{0}$ and $\ket{+}$. When this happens, the acceptance probability is $\frac{1}{2}$. 

Notice that the two containments above mean that under the commonly believed
complexity theoretical assumption,\footnote{Namely, the derandomization
assumption~\cite{KvM02,MV05} which states that: for some $\epsilon > 0$, some
language in ${\sf NE} \cap {\sf coNE}$ requires non-deterministic circuit of
size $2^{\epsilon n}$. \label{fn:derand}}  $\AM = \NP$, we have 
\AM{} = \StoqMA{} = \MA{} = \NP{}. Conjecture 
\ref{conj:stoqma-equals-ma} is, of course, weaker. 
We notice that this conjecture would also follow from another natural property of complexity classes, which
many classes, including \MA{}, \AM{}~\cite{FGMSZ89} and \class{QCMA}~\cite{JKNN12} possess: the property of being closed under perfect completeness 
(namely, that the class does not change if we require the verifier to accept yes-instances with probability $1$). By the results of \cite{BBT06,BT10}, 
if \StoqMA{} turns out to be closed under perfect completeness, this would imply it is equal to \MA{}.

\vspace{-0.5em}
\subsection{Results} 
Our main technical contribution concerns a problem that we call {\it projection uniform  stoquastic Local Hamiltonian}. This problem is a restricted variant of the stoquastic local Hamiltonian problem proposed by \cite{BBT06,BT10}. First, we require each local term to be a projection. Secondly, as in~\cite{AG19} (in the case of frustration-free Hamiltonians), we require that the terms are {\it uniform}; by this we mean that the groundspace of each local term is spanned by an orthonormal basis of quantum states, each being a {\it uniform} superposition of strings (see \Cref{def:uniformity}). 

Our main technical result then states that deciding whether a projection uniform  stoquastic Local Hamiltonian has ground-energy at most $\negl(n)$ or at least $1/\poly(n)$ (we denote this as the \PSLH{}$\left(\negl(n),1/\poly(n)\right)$ problem), is in \MA{}.

\begin{theorem}\label{lem:neglstoqma-ma}
\PSLH{}$\left(\negl(n),1/\poly(n)\right)$ is in \MA{}.
\end{theorem}

We argue in \Cref{lemma:SLH-is-StoqMA-negl-hard} that the $\StoqMA$-hardness proof of \cite{BBT06} can be easily patched to hold also with the restriction of the terms being projections and uniform; more precisely, their proof can be easily strengthened to show that
\PSLH{}$(\negl(n),1/\poly(n))$ is $\StoqMA(1-\negl(n),1-1/\poly(n))$-hard. Therefore, we have the following important implication of Theorem \ref{lem:neglstoqma-ma}: 

\begin{corollary}
$\StoqMA{}\left(1 - \negl(n),1-\frac{1}{\poly(n)}\right) \subseteq \MA.$
\end{corollary}

This result has a surprising consequence for the general  $\StoqMA(a,b)$ vs. $\MA$ question: it implies that
if we have error reduction for \StoqMA{} (in other words, if
Conjecture~\ref{conj:StoqMA-error-reduction} holds), then \StoqMA{} = \MA{}
(\Cref{fig:error-reduction})

\begin{corollary}[\StoqMA{} = \MA{}
is equivalent to error reduction for \StoqMA{}]\label{corr:equivalence}
Conjecture~\ref{conj:StoqMA-error-reduction} is true iff Conjecture~\ref{conj:stoqma-equals-ma} is true.
\end{corollary}

In fact, a weaker version of Conjecture \ref{conj:StoqMA-error-reduction}, 
in which the error is reduced not exponentially but rather just to being negligibly small (namely,
for any $a-b\geq 1/\poly(n), \StoqMA(a,b)) \subseteq \StoqMA\left(1-\negl(n),1-1/\poly(n)\right)$),
would already imply Conjecture \ref{conj:stoqma-equals-ma}. 
Interestingly, it suffices here that only the error in the completeness is reduced; this also appears in the proof of $\MA=\MA_1$ \cite{FGMSZ89}.

Optimistically, our result shows a possible path towards \StoqMA{} = \MA{}.
An analogous phenomenon holds for ${\sf QMA(2)}$:  
\cite{KMY03} showed that the ability to perform error reduction for ${\sf QMA(2)}$ would imply ${\sf QMA(2)} = {\sf QMA(k)}$, where $k$ indicates $k$ unentangled provers. In this case, it turned out to be a successful story: Harrow and Montanaro~\cite{HM13} indeed proved such an equivalence a few years later. 

With a more negative perspective, our result indicates that (dis)proving error reduction procedures for \StoqMA{} might be a difficult task.

\subsection{Implications of our results}

\begin{figure}[!ht]
     \centering
     \begin{subfigure}[b]{0.4\textwidth}
         \centering
         \includegraphics[width=\textwidth]{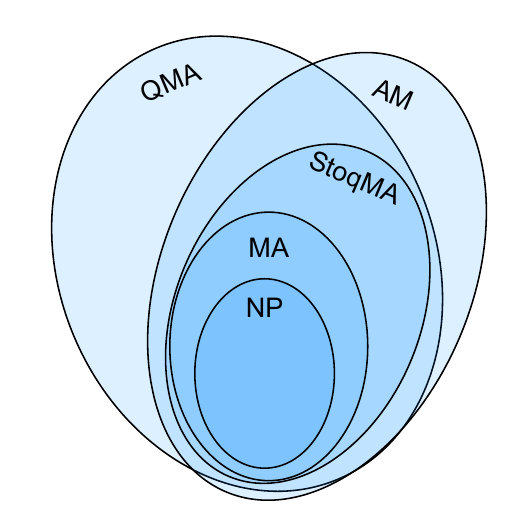}
         \caption{The currently known relation between the complexity classes.\\}
         \label{fig:known}
     \end{subfigure}
     \hfill
     \begin{subfigure}[b]{0.4\textwidth}
         \centering
         \includegraphics[width=\textwidth]{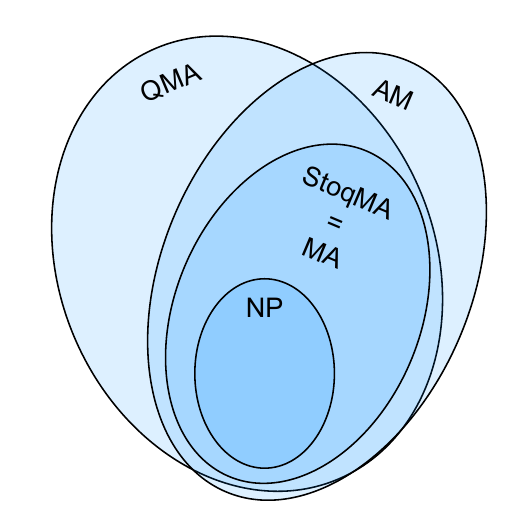}
         \caption{Relation between the complexity classes if error reduction is
         possible for \StoqMA.}
         \label{fig:error-reduction}
     \end{subfigure}
     ~\\
     \begin{subfigure}[b]{0.4\textwidth}
         \centering
         \includegraphics[width=\textwidth]{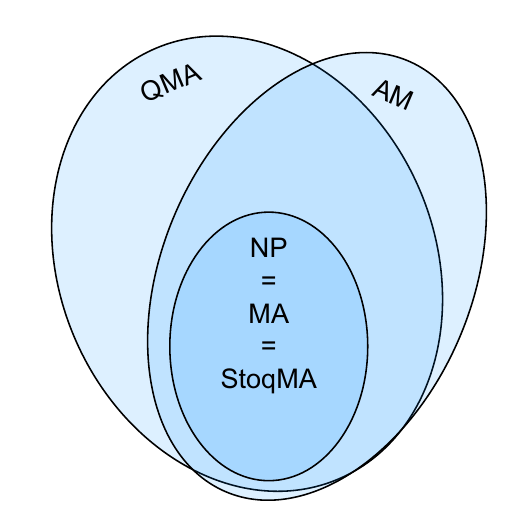}
         \caption{Relation between the complexity classes if error reduction is
         possible for \StoqMA{} and we assume weak derandomization assumptions.}
         \label{fig:error-reduction-derandomization}
     \end{subfigure}
     \hfill
     \begin{subfigure}[b]{0.4\textwidth}
         \centering
         \includegraphics[width=\textwidth]{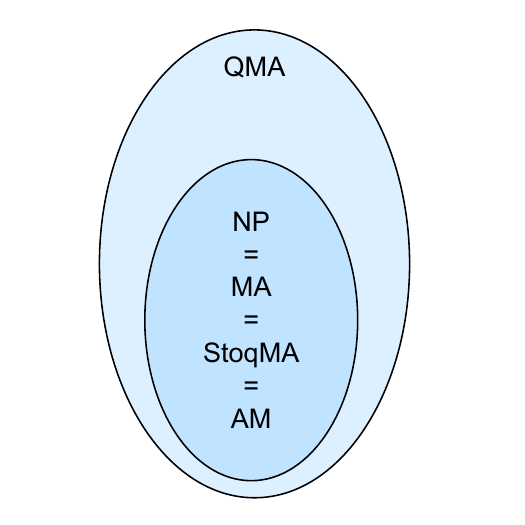}
         \caption{Relation between the complexity classes under strong
         derandomization assumptions.\\}
         \label{fig:strong-derandomization}
     \end{subfigure}
         \caption{Relation between the complexity classes \NP, \MA, \StoqMA,
         \AM{} and \QMA.}
        \label{fig:complexity-classes}
\end{figure}

We discuss now the implications of our result to the relation between the
complexity classes \NP, \MA, \StoqMA, \AM{} and \QMA, and to the corresponding
landscape of local Hamiltonians problems.  We first recall that the known
relation between these complexity classes is the following:
\[\NP \subseteq  \MA \subseteq \StoqMA \subseteq \AM \text{\quad and \quad} 
\NP \subseteq  \MA \subseteq \StoqMA \subseteq \QMA 
\text{\quad (\Cref{fig:known})}
.\]

However, under strong
derandomization assumptions (see 
\Cref{fn:derand}), we have that $\NP = \AM$ and therefore
\[\NP =  \MA  = \StoqMA = \AM  \subseteq \QMA
\text{\quad (\Cref{fig:strong-derandomization})}.\]
Notice that this assumption simplifies the landscape of the complexity of the
local Hamiltonian problem by collapsing the hierarchy proposed in
\cite{CM16}: the complexity of the local Hamiltonian problem with inverse
polynomial promise gap is either in \class{P}, \NP{}-complete or \QMA{}-complete.

The interesting consequences of our result apply when such strong
derandomization assumptions do not hold (or are not proven). More concretely,
we show that if error reduction for \StoqMA{} is proven
(Conjecture~\ref{conj:StoqMA-error-reduction}), then $\StoqMA{} = \MA$
(\Cref{fig:error-reduction}) and the four classes of local Hamiltonian problems
in \cite{CM16} become:  \class{P}, \NP{}-complete, \MA{}-complete or
\QMA{}-complete. We notice that \MA{} is a much more natural complexity class
than \StoqMA{}, so while this does not collapse any of these categories, it
qualitatively simplifies their landscape.

But if on top of Conjecture~\ref{conj:StoqMA-error-reduction}, we also assume
weak derandomization (i.e. $\MA = \NP$), then $\StoqMA = \NP$
(\Cref{fig:error-reduction-derandomization}), and we also have that the complexity of the local Hamiltonian problem with inverse
polynomial promise gap is either in \class{P}, \NP{}-complete or \QMA{}-complete.

\vspace{-0.5em}
\subsection{Proof overview}
The starting point of the proof of \Cref{lem:neglstoqma-ma} (given in Section \ref{sec:proof}) is the random walk used in \cite{BBT06,BT10}, to prove that the frustration-free stoquastic $k$-Local Hamiltonians problem is in \MA{}. 

More concretely, the input to the problem is a stoquastic local Hamiltonian $H = \frac{1}{m} \sum_{i \in [m]} H_i$, where each $H_i$ has norm at most $1$, and acts on at most $k$ out of an $n$ qubit system. Then, we have to decide if there exists a quantum state $\ket{\psi}$ such that $\bra{\psi}H\ket{\psi} = 0$, in which case we say that $H$ is frustration-free, or for all possible quantum states $\ket{\psi}$, $\bra{\psi}H\ket{\psi} \geq \frac{1}{\poly(n)}$.

 To prove containment in \MA{}, \cite{BBT06,BT10} define a random walk associated with $H$, which can be viewed as a walk on the following exponential-size (multi)graph $G(H)$. The vertices in the graph are all the possible $n$-bit strings, and the edges are defined as follows: for each  $i \in [m]$, we connect the two nodes (strings) $x,y$ iff $\bra{x}H_i\ket{y} \ne 0$.
\cite{BBT06,BT10} also define the notion of a {\it bad string},
which is a string that does not appear in the support of any of the groundstates
of some local term, i.e., $x$ is bad if there exists some $i$ such that $\bra{x}H_i\ket{x} = 0$ (and as expected, if $x$ is not bad, we say it is good).

The \MA{}-verification consists of running a random walk on $G(H)$ for polynomially many steps, starting at an $n$-bit string $x_0$ provided by the prover and rejecting if at any point along the walk a bad string is encountered.\footnote{
Notice that such a walk can be implemented efficiently since we since we can compute the neighbors of each string in polynomial-time given $H$.}
\cite{BBT06,BT10} showed that if the stoquastic Hamiltonian is frustration-free, then the connected component of any string in the support of any groundstate of the Hamiltonian, does not contain bad strings, and so 
any walk on the above defined graph, which starts from some string in a groundstate, does not reach bad strings.
On the other hand, if the Hamiltonian is at least $\frac{1}{p(n)}$ frustrated,
for some polynomial $p$, then there exists some polynomial $q$ such that a
$q(n)$-step random walk starting from any initial string reaches a bad string
with high probability.

In this work, we will keep the soundness part of the proof of \cite{BT10}, showing that for negative instances, no matter what string you start with, a random walk will reach a bad string after polynomially many steps with constant probability. However, we need a stronger statement for completeness. In the case of perfect completeness, it was true that {\it any} string $x$ in the support of a groundstate, would be a good witness, namely a random walk starting at $x$ will never reach a bad string. Once we relax the perfect completeness requirement, this is no longer the case: for example the groundstate could contain a bad string $\hat x$, and any random walk starting at $\hat x$ rejects. However, we are able to show that if $H$ has {\it negligible} frustration, then there must exist {\it some} initial string $x_0$ such that the probability that a polynomial-time random  walk starting from $x_0$ reaches a bad string, is negligibly small. 

The main idea to prove this result is to use a central result of \cite{AG19}, connecting the amount of frustration of the Hamiltonian, to a certain quantity of {\it expansion} in the graph. More precisely, the result we use says that if the Hamiltonian has only negligible frustration, then there must exist a non-empty set of nodes $S$ in $G(H)$ (recall that nodes correspond to $n$-bit strings) such that $i)$ all strings in the set are good strings, and $ii)$ the set has negligible expansion. The existence of such a set is proven in \Cref{lemma:nice-set}.

Given such a set, we prove in \Cref{lem:compl-stoqma}  that there exists one node in the set from which bad strings are hard to reach. To show this, we apply results about {\em escaping probabilities} of random walks~\cite{GharanT12}: if we start on a uniformly random element of the weakly expanding set of good strings $S$, then the probability that we leave this set is negligibly small. We can then apply an averaging argument, to deduce  that there must exist {\it some} initial string $x_0$ with the same property that starting the random walk from it, the probability to escape the set within polynomially many times steps of the walk, is negligible; this proves our result.

\vspace{-0.5em}
\subsection{Conclusions and open questions}
In this work, we prove a simple result which somewhat unexpectedly connects two problems; one, the question of the relation between \MA{} and \StoqMA{}, 
and the other, the question of error reduction for \StoqMA{}. This connection might lead to a clarification of the peculiar class \StoqMA{}; hopefully, it can point at a path by which it can be proven that 
\StoqMA{}=\MA{}. 

Of course, the main open problem that is highlighted by our work is whether error reduction for \StoqMA{} is possible. 

We mention also a related interesting question: this is the study of \StoqMA{} (and the  stoquastic Local Hamiltonian problem) with constant gap between completeness and soundness. In particular, if this problem can be solved in \NP{}, as is the case for its frustration-free variant~\cite{AG19}, then there would be a stronger connection between the Quantum PCP conjecture and the derandomization conjecture than the one proved in \cite{AG19}: a slightly weaker gap amplification procedure for stoquastic Hamiltonians would already imply \MA{}=\NP{}.\footnote{More precisely,  Let us consider a gap amplification procedure $\phi$ that maps stoquastic Hamiltonians to stoquastic Hamiltonians such that:
$a)$ if the ground-energy of $H$ is $0$, then the ground-energy of $\phi(H)$ is $\negl(n)$; and $b)$
if the ground-energy of $H$ is at least $1/\poly(n)$, then the ground-energy of $\phi(H)$ is at least some constant $\eps$.
\cite{AG19} proves that if such $\phi$ exists, then $\MA = \NP$. The weaker version of the gap amplification that we mention would allow $\phi(H)$ ground-energy to be some constant $\eps'$, for $\eps' < \eps$, if $H$ is frustration-free. See \cite{AG19} for a more details on the gap amplification procedure.}

\vspace{-0.7em}
\section{Preliminaries}
\subsection{Notation}
For $n \in \mathbb{N}^+$, we denote $[n] = \{0,...,n-1\}$. We we say that $f(n) = \negl(n)$ if $f = o\left(\frac{1}{n^c}\right)$ for every constant $c$.

\vspace{-0.5em}
\subsection{Graph theory and random walks}
For some undirected graph $G = (V,E)$, and $v \in V$, let $d_G(v) = |\{u \in V : \{v,u\}
\in E \}|$ be the degree of $v$. 
For some set of vertices $S \subseteq V$, we
define the edge boundary of $S$ as
$\partial_G(S) = \{\{u,v\} \in E: u \in S, v \not\in S\}$, the neighbors of $S$
as $N_G(S) = \{u \in \overline{S} : v \in S, \{u,v\} \in E\}$ 
(where $N_G(u)$ for some $u\in V$ is shorthand for $N_G(\{u\})$); and finally 
the volume of $S$
$vol_G(S) = |\{\{u,v\} : u \in S, v \in N_G(u) \}|$ and the
conductance of $S$, $\phi_G(S) = \frac{|\partial_G(S)|}{vol_G(S)}$. 

We define notions related to random walks on graphs; 
definitions 
are taken from \cite{LevinPeresWilmer2006}.
One step in the random walk on the graph $G$ starting on vertex $v$ is defined by moving to vertex $u$ with probability $p_{v,u} = \begin{cases}\frac{1}{d_G(v)}, & \text{if } \{u,v\}  \in E \\ 0 ,& \text{otherwise} \end{cases}$.

A {\it lazy} random walk stays put with probability half, and with probability half applies a step of the above random walk. 
A lazy random walk on a connected undirected graph always converges to a unique \textit{limiting distribution}; we denote this distribution over the vertices by $\pi$. 
It is well known that $\pi(v)$ is proportional to 
the degree of $v$, $d_G(v)$. 

We denote by $x_1,...,x_t \sim {\rm RW}(G,x_0,t)$ the distribution over $t$-tuple of variables, $x_1,...,x_t$, derived from running a $t$-step random walk on $G$ starting at $x_0$.

\begin{lemma}[Escaping probability, Proposition 3.1 in \cite{GharanT12}] \label{lem:escaping}
Let $G(V,E)$ be a graph. For any $S \subseteq V$ and integer $t$, we have that a $t$-step lazy random walk starting at a randomly chosen vertex $v \in S$, where $v$ is chosen with probability $\pi(v)$ restricted to $S$, leaves the set $S$ with expected probability at most $1-(1-\phi_G(S)/2)^t$. 
\end{lemma}

\vspace{-0.5em}
\subsection{Quantum computing}
We review now the concepts and notation of Quantum Computation that are used in
this work. We refer to Ref. \cite{NC02} for a detailed introduction of
these topics.

A pure quantum state of $n$ qubits is a unit vector in the Hilbert space
$\left\{\complex^{2}\right)^{\otimes n}$, where $\otimes$ is the Kroeneker (or tensor) product.
The basis for such Hilbert space is $\{\ket{i}\}_{i \in \{0,1\}^n}$.  
For some quantum state $\ket{\psi}$, we denote $\bra{\psi}$ as its conjugate transpose. The inner product between two vectors $\ket{\psi}$ and $\ket{\phi}$ is denoted by
$\innerprod{\psi}{\phi}$ and their outer product as
$\ketbra{\psi}{\phi}$.

We now introduce some notation which is somewhat less commonly used and more specific for this paper: the support of $\ket{\psi}$,  $supp(\ket{\psi}) = \{i \in \{0,1\}^n : \innerprod{\psi}{i} \ne 0\}$, is the set strings with non-zero amplitude.  
We call quantum state $\ket{\psi}$ {\it non-negative} if $\innerprod{i}{\psi}$ is real and $\innerprod{i}{\psi} \geq 0$ for
all $i \in \{0,1\}^n$.
For any $S \subseteq \{0,1\}^n$, we define the state $\ket{S} :=
\frac{1}{\sqrt{|S|}} \sum_{i \in S} \ket{i}$ as the subset-state corresponding to
the set $S$~\cite{Wat00}.

\vspace{-0.5em}
\subsection{Complexity classes}\label{sec:complexity-classes}
A (promise) problem $\calL = (\calL_{\rm yes}, \calL_{\rm no})$ consists of two non-overlapping subsets $\calL_{\rm yes}, \calL_{\rm no} \subseteq \{0,1\}^{*}$. 
We hereby define the main complexity classes that are considered in this work. 

\begin{definition}[\MA{}]
\label{def:MA}
A promise problem $\calL = (\calL_{\rm yes}, \calL_{\rm no}) \in$ \MA{} iff there exist a polynomial $p(n)$ and a probabilistic polynomial-time verifier $V(x,w)$ where $V$ takes as input a string $x \in \{0,1\}^*$ and a $p(n)$-bit witness $w$, and decides on acceptance or rejection of $x$ such that 
\begin{description}
    \item[Completeness.] If $x\in \calL_{\rm yes}$, then there exists $w$ such that $\Pr{V \text{ accepts } (x,w)} \ge 2/3$; 
    \item[Soundness.] If $x\in \calL_{\rm no}$, for any witness $w$, $\Pr{V \text{ accepts } (x,w)} \le 1/3$. 
\end{description}
\end{definition}

It is worth mentioning that \cite{BDOT06} provides an alternative
quantum-mechanical definition of $\MA$, namely, the verifier is a polynomial-size quantum circuit which uses only classical reversible gates (NOT, CNOT, Toffoli gate). This circuit receives a quantum witness, on top of auxiliary qubits, each of which is initialized either to the $\ket{0}$ or the $\ket{+}$ state. The verifier applies on this state the circuit, and then measures the output qubit in the computational basis and decides on acceptance/rejection.

By replacing the computational basis of the final measurement in the above definition, with a measurement in the Hadamard basis (arriving at the so-called \textit{stoquastic verifier}), this  leads to the complexity class $\StoqMA$: 

\begin{definition}[\StoqMA{}$(a,b)$, adapted from \cite{BBT06}]
\label{def:StoqMA}
A stoquastic verifier is a tuple $V=(n,n_w,n_0,n_+,U)$, where $n$ is the number of input bits, $n_w$ the number of input witness qubits, $n_0$ the number of input auxiliary qubits $\ket{0}$, $n_+$ the number of input auxiliary qubits $\ket{+}$ and $U$ is a quantum circuit on $n+n_w+n_0+n_+$ qubits with NOT, CNOT, and Toffoli gates. 
The acceptance probability of a stoquastic verifier $V$ on input string $x\in \{0,1\}^n$ and $n_w$-qubit witness state $\ket{\psi}$ is defined as $\Pr{V \text{  accepts } (x,\ket{\psi})}= \bra{\psi_{\rm in}}U^\dag\, \Pi_{\rm out}\, U \ket{\psi_{\rm in}}$. 
Here $\ket{\psi_{\rm in}} = \ket{x}\otimes \ket{\psi} \otimes \ket{0}^{\otimes n_0}\otimes \ket{+}^{\otimes n_+}$ is the initial state and $\Pi_{\rm out}=\ketbra{+}{+}_1\otimes I_{\rm else}$ projects the first qubit onto the state $\ket{+}$.\\
A promise problem $\calL=(\calL_{\rm yes}, \calL_{\rm no}) \in $ \StoqMA{}(a,b) iff there exists a
uniform family of stoquastic verifiers, such that for any fixed
number of input bits $n$ the corresponding verifier $V$ uses at most $p_1(n)$ qubits, $p_2(n)$ gates where both $p_1$ and $p_2$ are polynomials; and obeys the following: 
\begin{description}
    \item[Completeness. ] If $x \in \calL_{\rm yes}$, then there is some $\ket{\psi}$ s.t. $\Pr{V \text{ accepts } (x,\ket{\psi})} \geq a$; 
    \item[Soundness. ] If $x \in \calL_{\rm no}$, then for any $\ket{\psi}$, $\Pr{V \text{ accepts } (x,\ket{\psi})} \leq b$. 
\end{description}
\end{definition}

\begin{remark}\label{rem:soundness-bound}
We notice that we can assume $b \ge \frac{1}{2}$, WLOG: 
Notice that for a quantum circuit composed solely by classical reversible gates, if its input state is non-negative, so is its output, since classical gates do not modify the phases of the quantum state.
In this case, the final state of the $\StoqMA$ verifier's circuit on any non-negative witness is $\alpha_0 \ket{0}\ket{\psi_0} + \alpha_1 \ket{1}\ket{\psi_1}$, where $\alpha_b \in \mathbb{R}$ and $\ket{\psi_b}$ is a non-negative state and therefore, the acceptance probability is at least 
\[ 
\frac{1}{2}\left(\alpha_0^2 \innerprod{\psi_0}{\psi_0} + 
2\alpha_0 \alpha_1 \innerprod{\psi_0}{\psi_1} + 
\alpha_1^2 \innerprod{\psi_1}{\psi_1}\right) \geq
\frac{1}{2}\left(\alpha_0^2  + 
\alpha_1^2 \right) = \frac{1}{2}.
\]
\end{remark}

\begin{remark}\label{rem:straightforward-error-reduction}
As mentioned in \Cref{sec:intro}, error reduction for $\StoqMA(a,b)$ remains an open question since majority voting does not work (at least not in a straightforward way). To see this, notice that in order to perform such a majority voting, one would need to compute $\ell$ parallel runs of the original $\StoqMA$ verifier, measure the output qubit of each one of such runs in the Hadamard basis, and then compute the majority of the outcomes. But, unfortunately, this requires $\ell$ measurements in the Hadamard basis, whereas we are only allowed to perform a single one of such measurements by the definition of $\StoqMA$ verifiers.
\end{remark}

\vspace{-0.5em}
\subsection{Stoquastic Hamiltonians}
\label{subsec:stoq-LH}

The goal of this section is to describe the projection uniform  stoquastic Local Hamiltonian problem. We start with the definition of local Hamiltonians.

\begin{definition}[$k$-Local Hamiltonian]
A {\em Hamiltonian} on $n$ qubits is a Hermitian operator on $\complex^{2^n}$, namely, a complex Hermitian matrix of  dimension $2^n \times  2^n$. 
A Hamiltonian on $n$ qubits is called {\em $k$-Local} if it can be written as $H = \sum_{i = 1}^m H_i$, where each $H_i$ can be written in the form $H_i=\tilde{H}_i\otimes I$, where $\tilde{H}_i$ acts on at most $k$ out of the $n$ qubits.  
\end{definition}

We remark that WLOG we assume that each term $H_i$ is positive semi-definite since we could just add a constant $cI$ to the term to make it PSD; this only causes a constant shift in the eigenstates of the total Hamiltonian, hence does not change the nature of the 
problems we discuss here. 
We describe now the properties of the Hamiltonian in which we are interested. 

\begin{definition}[Ground-energy, frustration]\label{def:frustration}
The ground-energy of a Hamiltonian $H$ acting on $n$ qubits is defined as $\min_{\ket{\psi}} \bra{\psi}H\ket{\psi}$, where the minimum is over all possible $n$-qubit states (of norm $1$), a groundstate of $H$ is a state which achieves such a minimum, i.e., $\textup{argmin}_{\ket{\psi}} \bra{\psi}H\ket{\psi}$, and the groundspace of $H$ is the subspace spanned by its groundstates.
We say that a $H$ is $\eps$-frustrated if its ground-energy is at least $\eps$ and we say that $H$ is frustration-free if its ground-energy is $0$.
\end{definition}

We restrict ourselves to stoquastic Hamiltonians, which lie in between classical (diagonal) Hamiltonians and general quantum ones. 

\begin{definition}[Stoquastic Hamiltonian~\cite{BDOT06}]
 A $k$-Local Hamiltonian $H = \sum_{i = 1}^m H_i$ is called {\em stoquastic} in the computational basis if for all $i$, the off-diagonal elements of $H_i$ (the local terms) in this basis are non-positive\footnote{Klassen and Terhal~\cite{KT19} have a different nomenclature. They call a matrix Z-symmetric if 
the off-diagonal elements of the local terms are non-positive and they call a Hamiltonian stoquastic if all local terms can be made Z-symmetric by local rotations.
}.
\end{definition}

Notice that the definition of stoquasticity is basis dependent, and in our definition we assume that the stoquastic Hamiltonian is {\it given} in the basis where each of the {\it local} terms is stoquastic, i.e., has non-positive off-diagonal elements.

A property of a stoquastic local Hamiltonian is that the groundspace of the local terms can be decomposed as a sum of orthogonal non-negative rank-$1$ projectors.
\begin{lemma}[Groundspace of stoquastic Hamiltonians, Proposition 4.1 of \cite{BT10}]
\label{lem:gspace-structure}
Let $H$ be a stoquastic Hamiltonian and let $P$ be the projector onto its groundspace. It follows that 
\begin{align}
\label{eq:projector-groundspace}
P = \sum_j \kb{\phi_j}, 
\end{align}
where for all $j$, $\ket{\phi_j}$ is non-negative and for $j \ne j'$,
$\innerprod{\phi_{j'}}{\phi_j} = 0$.
\end{lemma}

\vspace{-0.5em}
\subsubsection{Projection Uniform Stoquastic Hamiltonians}
We now restrict ourselves to stoquastic Local Hamiltonians whose local terms have groundspaces which are much simpler, namely, the groundspace is spanned by orthogonal subset-states. In addition, we also require each $H_i$ to be itself a projection.

\vspace{-0.2em}
\begin{definition}[projection uniform stoquastic Local Hamiltonian]\label{def:uniformity}
A stoquastic Local Hamiltonian $H = \frac{1}{m} \sum_{i = 1}^m H_i$ is called projection uniform if for all $i$: $a)$ $H_i$ is a projection and $b)$ when considering the groundspace projector $P_i = I - H_i$, there exist disjoint sets $S_{i,j} \subseteq \{0,1\}^n$ such that 
$P_i = \sum_{j} \kb{S_{i,j}}$,
namely, the groundspace can be spanned by subset states on the disjoint sets
  $S_{i,j}$.\footnote{Notice that while there are multiple ways of decomposing
  the groundspace of $H_i$, there is at most one way of doing it if we require
  that all
  vectors in the decomposition are non-negative. In this case, the partition
  $\{S_{i,j}\}_j$ is unique.  }
\end{definition}

We can finally define the projection uniform stoquastic $k$-Local Hamiltonian problem.

\vspace{-0.2em}
\defproblem{projection uniform stoquastic $k$-Local Hamiltonian problem (\PSLH($\alpha,\beta$)}{def:local-hamiltonian}{We are given three parameters: 
   $k \in \mathbb{N}^*$ which is called the locality, 
   $\alpha,\beta :  \mathbb{N} \rightarrow [0,1]$ which are a non-decreasing functions with $\alpha(n) < \beta(n)$.The {\em projection uniform stoquastic  $k$-Local Hamiltonian} problem with completeness $\alpha$ and soundness $\beta$, and with parameters $k,\eps$, is 
   the following promise problem. Let $n$ be the number of qubits of a quantum system.
  The input is a set of $m(n)$ uniform stoquastic Hamiltonians $H_1, \ldots, H_{m(n)}$
  where $m$ is a polynomial, s.t. $\forall i \in [m(n)] :$ $H_i$ is a projection which acts on $k$ qubits out of the $n$ qubit system. 
Let $H = \frac{1}{m} \sum_{i = 1}^m H_i$. We are promised that one of the following two conditions hold, and we need to decide which one:
}
{There exists a $n$-qubit quantum state
       $\ket{\psi}$ such that
      $\bra{\psi} H \ket{\psi}
        \leq \alpha$}
{
For all $n$-qubit quantum states $\ket{\psi}$
      it holds that
      $\bra{\psi} H \ket{\psi}
        \geq \beta.$
        }

It turns out that considering this restricted family of Hamiltonians does not harm the \StoqMA{} hardness: 
\vspace{-0.2em}
\begin{lemma}[\PSLH{}$(\negl,1/\poly)$ is \StoqMA{}$(1-\negl,1-1/\poly)$-hard]
\label{lemma:SLH-is-StoqMA-negl-hard}
For any polynomial $p_1(n)$ and negligible function $\eps(n)$, there exists a polynomial $p_2(n)$ such that the problem  \PSLH{}$\left(\eps(n),\frac{1}{p_2(n)}\right)$ is \StoqMA{}$\left(1-\eps(n),1-\frac{1}{p_1(n)}\right)$-hard. 
\end{lemma}
\vspace{-0.5em}
\begin{proof}
The proof is analogous to the $\MA$-hardness proof of stoquastic $6$-SAT
  (Appendix B in \cite{BT10}), in which Kitaev's circuit to Hamiltonian
  construction is used to devise a local Hamiltonian whose ground state is the
  history of the verification computation. Given a \StoqMA{} verifier with
  completeness $1-\eps(n)$ and soundness $1-\frac{1}{p_1(n)}$, we end up with a
  projection uniform stoquastic Hamiltonian $H$ by following the construction in
  \cite{BT10}, except we replace the output constraint
  $\Pi_{out}:=\ketbra{0}{0}_1$ in \cite{BT10} by $\ketbra{+}{+}_1$. More concretely, for a \StoqMA{} circuit consisting of $T$ gates $G_1,...,G_T$ acting on a $P$-qubit proof, $A$ auxiliary qubits $\ket{0}$ and $R$ auxiliary qubits $\ket{+}$, we consider the terms:
  \begin{align*} 
  H^{aux0}_{i} &= \kb{01}_{1,T+P+i}, \forall i \in [A] \\
  H^{aux+}_{i} &= \kb{0-}_{1,T+P+A+i}, \forall i \in [R] \\
H^{clock}_{t} &= \kb{01}_{t,t+1}, \forall t \in [T]\\
H^{prop}_{t} &=  I - \sum_{z} \left(\ket{100}\ket{z} + \ket{110}G_t\ket{z}\right) 
\left(\bra{100}\bra{z} + \bra{110}\bra{z}G_t\right)_{t-1,t,t+1,W(t)}, \forall t \in [T] \\ 
H^{out} &= \kb{1-}_{T,T+1}, \end{align*}
where $W(t)$ is the set of qubits on which the $t$-th gate acts. Notice that each of these terms are projections, that $H^{aux0}_i$, $H^{aux+}_i$,  and $H^{clock}_t$ are diagonal, and that all of the off-diagonal elements of $H^{prop}_{t}$ and $H^{out}$ are non-negative. Therefore, the sum of all of these terms is a projection uniform stoquastic Hamiltonian.

As is typically the case in the circuit to Hamiltonian construction proofs, the completeness parameter is preserved, namely, the ground-energy of $H$ is upper-bounded by $\eps(n)$ - this is because the history state $\ket{\eta}$ for the witness which is rejected with probability at most $\eps(n)$, satisfies 
$\bra{\eta}H_{out}\ket{\eta}= \|H_{out}\ket{\eta}\|^2\le \eps(n)$.

For soundness, since the \StoqMA{} verifier accepts any witness state with probability at most $1-1/p_1(n)$, the statement at the beginning of Section 14.4.4 in \cite{KSV02} (with $\epsilon$ in \cite{KSV02} being equal to $1-1/p_1(n)$) implies that the ground-energy of $H$ is lower-bounded by 
\[ \lambda_{\rm min}(H) \geq \frac{C_0 \left(1-\sqrt{1-1/p_1(n)}\right)}{L^3} \geq
\frac{C_0 \left(1-(1-1/2p_1(n))\right)}{L^3}
= \frac{C_0}{2 p_1(n) L^3}
:= \frac{1}{p_2(n)}, \]
where $C_0$ is some positive constant, $L$ is the number of clock qubits which is a polynomial, and $p_2(n)$ is clearly a polynomial. It completes the proof. 
\end{proof}

\section{Proof of main result}\label{sec:proof}

In this section we prove \Cref{lem:neglstoqma-ma}. To prove that \PSLH{}($\negl,1/\poly$) is in \MA{}, it suffices to show that there exists a negligible function $\eps$ and some value $t = \poly(n)$ such that: 
\begin{description}
    \item[Completeness.] If $H$ is a \textit{yes}-instance, then there exists a string $x$ such that any $t$-step random walk starting at $x$ will reach a bad string with probability at most $\eps(n)$; 
	\item[Soundness.] If $H$ is a \textit{no}-instance, then all $t$-step random walk will reach a bad string with constant probability. 
\end{description}

We need to find $t$ which satisfies both conditions. 
In \Cref{lem:sound-bt}, which is proven in \Cref{sec:BT}, we derive the soundness: we show that there exists some $t  = \poly(n)$ which satisfies the soundness condition. 
This part in fact follows almost exactly the proof of \cite{BT10}. 
We will later (in \Cref{sec:proof-main,sec:nice-sets}) show that the completeness condition is satisfied for {\em any} polynomial $t$, which completes the proof of 
\Cref{lem:neglstoqma-ma}.

\vspace{-0.5em}
\subsection{Soundness: Recalling the random walk argument from \cite{BT10}}\label{sec:BT}

We describe now the random walk approach of \cite{BBT06,BT10} to show that the stoquastic frustration-free Hamiltonian problem is in \MA{}. As in \cite{AG19}, we restrict ourselves to the simplified version of their result that works for projection uniform stoquastic Hamiltonian, which is sufficient for our purposes given that this problem is $\StoqMA$-hard as proven in \Cref{lemma:SLH-is-StoqMA-negl-hard}. 
We start with defining the configuration graph for a given projection uniform stoquastic Hamiltonian. 

\begin{definition}[Graph from projection uniform stoquastic Hamiltonian]
\label{def:configuration-graph}
  Let $H = \frac{1}{m}\sum_{i} H_i$  be a projection uniform stoquastic $k$-local Hamiltonian on $n$ qubits. We define the undirected multi-graph $G(H) = (\{0,1\}^n, E)$ where $E$ is defined as follows.
  Let $M = (2^k)!$ and for each $i \in [m]$, let $S_i^x = \{y : \bra{x}P_i\ket{y} > 0\} \cup \{x\}$.\footnote{The forced inclusion of $x$ in $S_i^x$ guarantees that all vertices have self-loops and it does not change the random walk because, as we will see later, if $\bra{x}P_i\ket{x} = 0$, the random walk halts.} For each $i \in [m]$, if $S_i^x \ne \emptyset$, for each $y \in S^x_i$ we add $\frac{M}{|S_i^x|}$ edges between $(x,y)$. 
\end{definition}

We now argue that the lazy random walk~\footnote{We use the term lazy
random walk to highlight the fact that in \Cref{def:configuration-graph}, we
added self-loops for every node. We notice that since $x \in S^x_i$ for every
$i$ and $1 \leq |S^x_i| \leq k$, this is equivalent to the
standard notion of lazy random walk where we stay in the same vertex with
constant probability $\frac{1}{k}$.} on the above configuration graph is the same
walk as used in \cite{BBT06,BT10}.  The random walk in
\cite{BBT06,BT10} is defined in the following way: first we pick a term $H_i$
uniformly at random out of the $m$ terms, and then we move from $x$ to a random
$y$ such that  $\bra{y}H_i\ket{x} \ne 0$ (we say in this case that $y$ and $x$
are neighbors via $H_i$). As usual, we consider the lazy version of this walk.
We note that the number of edges leaving a node $x$ is 
\[\sum_{i \in [m]} |S_i^x| \cdot \frac{M}{|S^x_i|} = m M, \] and therefore the
probability to move  the string $x$ to a string $y$ in this walk is 
\[
  \frac{M/|S_i^x|}{m M} = 
  \frac{1}{m|S_i^x|}, 
\] summed over all $i$'s such that $x,y$ are neighbors via $H_i$. This gives the same ratio between probabilities to move to different $y$'s from a given $x$, 
as in the lazy simple random walk induced on the graph in  \Cref{def:configuration-graph}, and thus we have an equivalent lazy random walk.  

We note that in order to use known results in the literature regarding simple random walks, such as Lemma \ref{lem:escaping} we take from \cite{GharanT12}, we prefer to stick to the multi-graph notation and we add multiple edges that correspond to the weight of the random walk described in \cite{BBT06,BT10}. 

\cite{BBT06,BT10} also define a notion of bad strings, which are strings that do not belong to any groundstate of some of the local terms.
\begin{definition}[Bad strings \cite{BT10}]\label{def:bad-string}
Given a local Hamiltonian $H=\frac{1}{m}\sum_{i} H_i$, we say that $x$ is {\it
  bad} for $H$ if there exists some $i \in [m]$ such that $\bra{x}H_i\ket{x} = 0$. 
\end{definition}

The random walk starts from some initial string $x_0$ sent by the prover, runs for $T$ steps on this graph and at the end accepts iff no bad string was encountered in the walk.
We remark that given some string $x$, one can compute in polynomial time all neighbors of $x$ in $G(H)$ as well as decide whether $x$ is bad for $H$, by just inspecting the groundspace of each local term.

\begin{figure}[H]
\rule[1ex]{\textwidth}{0.5pt}
\vspace{-20pt}
\begin{enumerate}
 \item Let $x_0$ be the initial string.
 \item Repeat for $t=0,...,T-1$:
 \begin{enumerate} 
   \item If $x_t$ is bad for $H$, reject
   \item Pick $i \in [m]$ uniformly at random
   \item Pick a neighbor $x_{t+1}$ of $x_t$ via $H_i$ uniformly at random
 \end{enumerate}
 \item Accept
\end{enumerate}
\vspace{-.2cm}
\rule[2ex]{\textwidth}{0.5pt}\vspace{-.5cm}
\caption{Random Walk from \cite{BT10} (simplified to the uniform case)}
\label{algo:bt}
\end{figure}

\begin{lemma}[Soundness, adapted from Section $6.2$ of \cite{BT10}]\label{lem:sound-bt}
For every polynomial $p$, there exists some polynomial $q$ such that if $H$ is at least $\frac{1}{p(n)}$-frustrated, then for every string $x$, for $T=q(n)$, the random walk from \Cref{algo:bt} reaches a bad string (and thus rejects) with constant probability.
\end{lemma}

This completes the soundness part. 
The intuition of the proof is easier to explain in the projection uniform case: as noticed by \cite{AG20}, in this case, the frustration of a uniform stoquastic Hamiltonian $H$ is related to the expansion of the graph $G(H)$. Therefore, with inverse polynomial lower-bounds on the expansion of the graph and on the fraction of bad strings, one can show that there exists a polynomial $q$ such that a $q(n)$-step random walk on $G(H)$ starting at any vertex must encounter a bad string with high probability.

\vspace{-0.5em}
\subsection{Negligible frustration implies weakly expanding set of good strings}\label{sec:proof-main}

In \cite{BT10}, completeness was proven for the case of frustration-free Hamiltonian. 
\begin{lemma}[Completeness, adapted from Section $6.1$ of \cite{BT10}]\label{lem:bt-completeness}
If $H$ is frustration-free, then there exists some string $x$ such that there are no bad-strings in the connected component of $x$.
\end{lemma}

To prove this, \cite{BBT06,BT10} show that if $H$ is frustration-free, the set of strings in the support of any of its groundstates, form a union of  connected components of $G(H)$. Since $H$ is frustration-free, any such connected component must contain only good strings, since bad strings exhibit non-zero energy. Therefore any path starting at a vertex in this connected component never finds a bad string, and the verifier will always accept. 

Here, we do not have frustration-freeness, so we have to work harder. Due to the negligible frustration in the yes case, we have the following "approximation" of the situation of connected components of good strings 
(the next subsection uses this lemma to prove 
completeness, by a random walk argument): 

\begin{lemma}[Weakly expanding set of good strings]\label{lemma:nice-set}
For every negligible function $\eps$, there exists a negligible function $\eps'$ such that the following holds.
If a projection uniform stoquastic $k$-local Hamiltonian $H$ has ground-energy at most $\eps(n)$, then there exists some set $S \subseteq \{0,1\}^n$ of good strings such that $|\partial_{G(H)}(S)| < \eps'(n) |S|$.
\end{lemma}

In order to prove \Cref{lemma:nice-set}, we first list two auxiliary lemmas from \cite{AG19}.

\begin{lemma}[Projection on boundary lower bounds energy - Lemma $5.2$ from \cite{AG19}]\label{lem:connection-frustration-missing-strings}
Let $H$ be a projection uniform stoquastic Hamiltonian  and $\ket{\psi} = \sum_{x} \alpha_x \ket{x}$ be a positive state. Let $N$ be the set  $\{x \in  supp(\ket{\psi} : \exists i \in [m], y \not\in supp(\ket{\psi}), \bra{x}P_i\ket{y} > 0\}$. Then $\bra{\psi}H\ket{\psi} \geq \frac{1}{2^k m} \sum_{x \in N} \alpha_x^2$.
\end{lemma}

To get a rough intuition behind this lemma, we consider two strings $x \in supp(\ket{\psi})$ and $y  \not\in supp(\ket{\psi})$ that are neighbors via some term $H_i$. The contribution of $y$ to $\bra{\psi}H_i\ket{\psi}$ is, on one hand, at least (some function of) $\alpha_x$, the amplitude of $x$ in $\ket{\psi}$, and, on the other hand, we expect a factor of $\frac{1}{2^k}$ to appear, corresponding to the fact that if $\bra{y}P_i\ket{x}>0$ it must be at least $1/2^k$ since the subset states comprising the groundspace of $H_i$ have at most $2^k$ strings in their support. 

\begin{lemma}[Existence of a "nice" low energy state - Lemma $5.3$ of \cite{AG19}] \label{lemma:structure-gstate-negligible}
Let $H$ be a projection uniform  stoquastic Hamiltonian with non-zero ground-energy at most $\frac{1}{f(n)}$. Then for any $g(n)>0$, there exists some non-negative state $\ket{\psi}$ that does not contain bad strings for $H$, contains only strings with amplitude at least $\delta = \frac{1}{\sqrt{g(n)|S|}}$ and has frustration at most $\frac{1}{f(n)(1 - \frac{m}{f(n)} - \frac{1}{g(n)} )}$, where $S = supp(\ket{\psi})$ .
\end{lemma}

The proof of this lemma follows by starting from the true groundstate of $H$, and then removing all bad strings and strings with amplitude smaller than $\delta$. The result follows by a simple calculation of the frustration of the remaining (renormalized) state.
With these lemmas, we can prove \Cref{lemma:nice-set}
\begin{proof}[Proof of \Cref{lemma:nice-set}]
Let $\ket{\psi}$ be the state that comes from \Cref{lemma:structure-gstate-negligible}, by picking $f(n) = 1/\eps(n)$ and $g(n) = 1/\sqrt{\eps(n})$. As stated by \Cref{lemma:structure-gstate-negligible}, $\ket{\psi}$ does not contain bad strings, all the amplitudes are at least $\delta = \sqrt{\frac{\sqrt{\eps(n)}}{|S|}}$, where $S = supp(\ket{\psi})$, and this state has frustration 
\begin{align}\label{eq:bound-frustration}
    \bra{\psi}H\ket{\psi} < \frac{\eps(n)}{1 - m\eps(n)-\sqrt{\eps(n)}} \leq 2\eps(n).
\end{align} 
As in Lemma \ref{lem:connection-frustration-missing-strings}, let us define $N = \{ x \in S: \exists y \not\in S, i \in [m] \text{ s.t. } \bra{x}P_i\ket{y} > 0 \}$.
We first show that $|N| \leq m 2^{k+1} \sqrt{\eps(n)} \cdot |S|$. 
Let us assume towards contradiction that 
\begin{align}\label{eq:assumption-proof}
    \frac{|N|}{|S|} > m 2^{k+1} \sqrt{\eps(n)}.
\end{align}
We have that
\begin{align*}
   2\eps(n) > \bra{\psi}H\ket{\psi} \geq \frac{1}{m2^k} \sum_{x \in N} \alpha_x^2 \geq \frac{|N| \delta^2}{m 2^k} > 
   2\delta^2 \sqrt{\eps(n)}|S|=
   2\eps(n),
\end{align*}
where the first inequality holds from \Cref{eq:bound-frustration}, the second inequality holds from \Cref{lem:connection-frustration-missing-strings}, the third inequality comes from the fact that $\alpha_x \geq \delta$, the fourth inequality comes from the assumption on $|N|$ and the last equality comes from the definition of $\delta$.
This is a contradiction and therefore \Cref{eq:assumption-proof} does not hold.

We now notice that $N$ counts vertices in the boundary of $S$ in $G(H)$, whereas to finish the proof of Lemma \ref{lemma:nice-set}, we need to upper bound on the number of edges. By the definition of $G(H)$ (\Cref{def:configuration-graph}), each vertex is connected to at most $m2^k(2^k)!$ edges. Therefore, the result follows from the negation of \Cref{eq:assumption-proof} by choosing $\eps'(n) = m^22^{2k+1}(2^{k})! \sqrt{\eps(n)}$.
\end{proof}

\vspace{-0.8em}
\subsection{The existence of a witness in the negligible frustration case} \label{sec:nice-sets}
Using \Cref{lemma:nice-set}, we show that if a stoquastic Hamiltonian is negligibly frustrated, then there exists a string which is a ``good starting point'' for the random walk, i.e., there exists a vertex $x^*$ such that for any polynomial $p$, a $p(n)$-step random walk starting from $x^*$ finds a bad string with negligible probability. 

\begin{lemma}\label{lem:compl-stoqma}
For every negligible function $\eps$, there exists a negligible function $\hat\eps(n)$ such that the following holds.
If a projection uniform  stoquastic $k$-local Hamiltonian $H$ has ground-energy at most $\eps(n)$, then there exists a string $x_0 \in \{0,1\}^n$ such that for any polynomial $p$, the probability that a $p(n)$-step -random-walk starting at $x_0$ finds a bad string is bounded from above by $\hat\eps(n)$. 
\end{lemma}
\vspace{-0.5em}
\begin{proof}
From \Cref{lemma:nice-set}, there exists some set $S \subseteq \{0,1\}^n$ of good strings 
such that $|\partial_{G(H)}(S)| < \eps'(n) |S|$, for some $\eps'(n) = \negl(n)$.

In particular, this implies that $\phi(S) = \frac{|\partial_{G(H)}(S)|}{vol(S)} \leq \frac{|\partial_{G(H)}(S)|}{|S|}  < \eps'(n)$, where in the first inequality we use the fact that every $x \in S$ has a neighbor (not necessarily in $S$).\footnote{Notice that we can assume this without loss of generality, since if any $x \in S$ (which implies that $x$ is good) is not connected to any other string, a random walk starting at it never finds a bad string since it stays stationary in a good string. In this case our statement is trivially true.}
 Therefore, if we pick $x_0 \in S$ at random with probability $\pi_S$ (namely the probability proportional to the limiting distribution restricted to $S$), and perform a $p(n)$-step random walk starting from $x_0 \sim \pi_S$, we have by \Cref{lem:escaping} that the expected probability of leaving the set $S$ during this walk is at most
\begin{align}\label{eq:averaging}
\mathop{\mathbb{E}}_{x_0 \sim \pi_S} \mathop{\mathrm{Pr}}_{x_1,...,x_t \sim {\rm RW}(G,x_0,t)} \left[ \exists i \in [t] \text{ s.t. } x_i \not\in S \right]\leq 1- (1-\eps'(n))^{p(n)} \leq  p(n) \eps'(n):= \hat\eps(n).\end{align}

Notice that if \Cref{eq:averaging} is true, then there must exist some value $x_0$  such that 
\begin{align}\mathop{\mathrm{Pr}}_{x_1,...,x_t \sim {\rm RW}(G,x_0,t)} \left[\exists i \in [t] \text{ s.t. } x_i \not\in S \right]\leq  \hat\eps(n),
\end{align}
by an averaging argument. Since all strings in $S$ are good, we have that a random walk starting at $x_0$ finds a bad string with probability at most $\hat\eps(n)$.
\end{proof}

We are now ready to prove \Cref{lem:neglstoqma-ma}.

\vspace{-0.2em}
\begin{proof}[Proof of \Cref{lem:neglstoqma-ma}]
As in \cite{BT10}, the verification algorithm consists of the random walk described in \Cref{algo:bt} for $T = q(n)$, where $q$ is the polynomial from \Cref{lem:sound-bt}.

Since $q$ is a polynomial, we have from \Cref{lem:compl-stoqma} that if $H$ is a \textit{yes}-instance of \PSLH{}$(\negl(n),$ $1/\poly(n))$, then there exists some $x_0$ such that the random walk rejects with probability at most $\hat\eps(n)$, for some negligible function $\hat\eps$.

From \Cref{lem:sound-bt}, we have that for any \textit{no}-instance $H$ of \PSLH{}$(\negl(n),1/\poly(n))$ and any initial string $x_0$, the probability that the random walk in \Cref{algo:bt} rejects is bounded from below by a constant $>0$. 
\end{proof}

\section*{Acknowledgements}
\noindent
D.A. and Y.L. were gratefully supported by ISF Grant No. 1721/17, and part of this work was done while Y.L. was affiliated with the Hebrew University of Jerusalem. 
Part of this work was done while A.B.G. was affiliated to CWI and QuSoft and part of it was done while A.B.G. was visiting
the Simons Institute for the Theory of Computing A.B.G. is supported by ANR JCJC TCS-NISQ ANR-22-CE47-0004, and by
the French National Research Agency award number ANR-22-PNCQ-0002.

\bibliographystyle{quantum}
\bibliography{StoqMA-negl-in-MA.bib}

\appendix

\end{document}